\documentclass[reprint,aps,prf,showpacs,floatfix,notitlepage,superscriptaddress]{revtex4-2}
\usepackage{amssymb,amsmath,bm,bbm}
\usepackage{graphicx}
\usepackage[colorlinks=true,allcolors=blue]{hyperref}

\begin{document}
\title{Minimum Dissipation Theorem for Microswimmers}
\author{Babak Nasouri}
\affiliation{Max Planck Institute for Dynamics and Self-Organization (MPIDS), 37077 G\"ottingen, Germany}
\author{Andrej Vilfan}
\email{andrej.vilfan@ds.mpg.de}
\affiliation{Max Planck Institute for Dynamics and Self-Organization (MPIDS), 37077 G\"ottingen, Germany}
\affiliation{Jo\v{z}ef Stefan Institute, 1000 Ljubljana, Slovenia}
\author{Ramin Golestanian}
\affiliation{Max Planck Institute for Dynamics and Self-Organization (MPIDS), 37077 G\"ottingen, Germany}
\affiliation{Rudolf Peierls Centre for Theoretical Physics, University of Oxford, Oxford OX1 3PU, United Kingdom}
\date{\today}

\begin{abstract}
  We derive a theorem for the lower bound on the energy dissipation
  rate by a rigid surface-driven active microswimmer of arbitrary
  shape in a fluid at low Reynolds number.  We show that, for any
  swimmer, the minimum dissipation at a given velocity can be
  expressed in terms of the resistance tensors of two passive bodies
  of the same shape with a no-slip and perfect-slip boundary. To
  achieve the absolute minimum dissipation, the optimal swimmer needs
  a surface velocity profile that corresponds to the flow around the
  perfect-slip body, and a propulsive force density that corresponds
  to the no-slip body.  Using this theorem, we propose an alternative
  definition of the energetic efficiency of microswimmers that, unlike
  the commonly-used Lighthill efficiency, can never exceed unity. We
  validate the theory by calculating the efficiency limits of
  spheroidal swimmers.
\end{abstract}

\maketitle

Microswimmers are natural or artificial self-propelled microscale objects moving through a fluid at low Reynolds numbers, such that viscous forces dominate over inertia \cite{Gompper2020}. 
The swimming motion can arise from periodic changes in the shape of the swimmer, which have to be non-reciprocal in time \cite{purcell1977,najafi2004,Nasouri.Golestanian2019}. Organisms propelled by bacterial or eukaryotic flagella rely on non-reciprocal shape changes of their flagella. Many other microorganisms are propelled by thousands of cilia that all beat in an asymmetric fashion. Their beating could in principle be described as a shape change, but it is usually more insightful to use a coarse-grained approach 
in which the cilia are replaced by a propulsive layer that generates an effective tangential slip velocity along the surface \cite{lighthill1952,blake1971,Blake1973,julicher2009,osterman2011,vilfan2012,Pedley.Goldstein2016,Zantop.Stark2020}. Most artificial microswimmers rely on the self-phoretic mechanism and are therefore driven by a slip velocity by design \cite{anderson91,golestanian2005,Ebbens2018,nasouri2020,pohnl2020}.

The energetic efficiency of microswimmers is commonly defined by Lighthill's criterion as the power, needed when an external force moves a swimmer with drag coefficient $R$ with a speed $V$, divided by the dissipated power $P$ when the self-propelled swimmer moves with the same speed, namely $\eta_\text{L}={R V^2}/{P}$ \cite{lighthill1952}.
Maximizing $\eta_\text{L}$ will always provide the minimum power needed to achieve a certain swimming speed, or, conversely, the maximum speed that can be achieved with a given power. However, $\eta_\text{L}$ is not an efficiency in the thermodynamic sense and it can, in principle, exceed 100\% \cite{Childress2012}. For instance, \citet{leshansky2007} showed that the Lighthill efficiency of a prolate spheroid diverges with the aspect ratio, becoming infinite for a thin needle. 
It is possible to introduce an efficiency that has an upper bound by evaluating the potential ability of the swimmer to tow a tethered cargo \cite{Raz.Leshansky2008,Childress2012}, although this definition will depend on the specific (geometric) features of the cargo \cite{Golestanian2008b}.

\begin{figure}[b]
\centering \includegraphics[width=0.9\columnwidth]{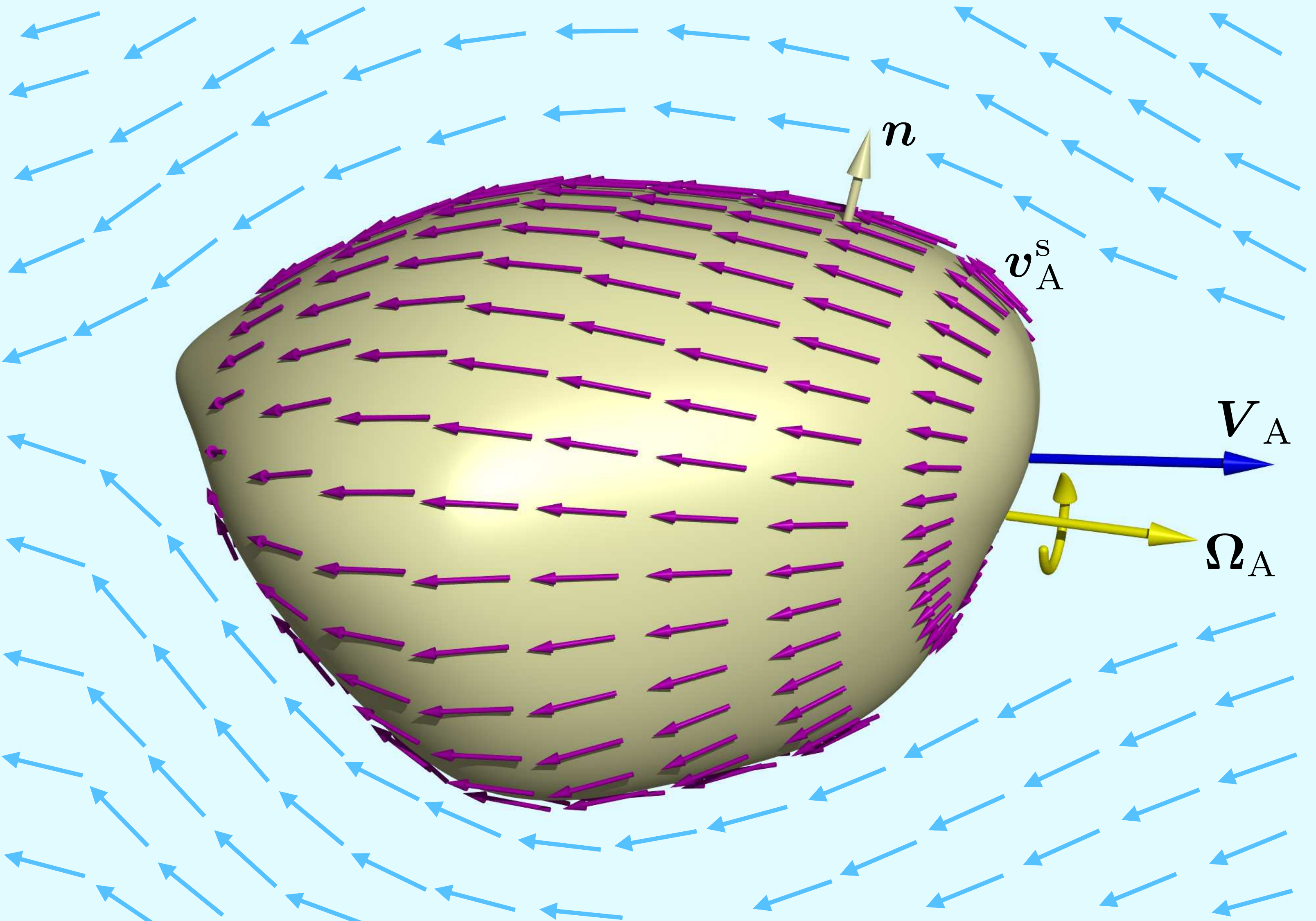}
\caption{A surface-slip driven swimmer of arbitrary shape with normal vector $\bm{n}$. The fluid at the boundary has a slip velocity $\bm{v}_\text{A}^\text{s}$ relative to the swimmer body. The swimmer moves with the rigid-body velocity $\bm{\mathsf{V}}_\text{A}=\left[\bm{V}_\text{A}~\bm{\Omega}_\text{A}\right]$, where $\bm{V}_\text{A}$ and $\bm{\Omega}_\text{A}$ are the translational and rotational velocities. The magenta and light-blue arrows show the schematic of the slip velocity and the streamlines, in the co-moving frame.}
  \label{fig:fig1}
\end{figure}

Microswimmers driven by an effective surface slip velocity have two contributions to the dissipation: external (in the \textit{outer} problem), due to the shearing motion of the surrounding fluid, and internal (in the \textit{inner} problem), due to losses in the propulsive layer. The latter, which focuses on the dissipation in the propulsive layer \cite{vilfan2012}, has been the focus of several studies on the grounds that this is often the dominant contribution, for example, in ciliated microorganisms \cite{Keller.Wu1977,Ito.Ishikawa2019}. 
The former, on the other hand, concerns the energy loss due to viscous dissipation, which has also been discussed analytically for spherical \cite{stone1996,michelin2010}, and spheroidal \cite{leshansky2007} swimmers, or computationally for more general axisymmetric swimmers \cite{guo2020}. Studies on phoretic swimmers also add the dissipation of the chemical mechanism that leads to the surface slip \cite{sabass2010,sabass2012}. Here, our focus is on the the external dissipation, which can set a fundamental limit on swimming efficiency, independent of the details of the driving mechanism. Since 1973, when John Blake proposed that minimum energy dissipation theorems be formulated \cite{blake_comment}, a number of analytical and numerical studies on efficiency limits of microswimmers have emerged. However, general statements on efficiency bounds have remained scarce.

 In this Letter, we propose a theorem that sets a fundamental lower bound on the external dissipation $P_\text{A}$ around a surface-driven active microswimmer of any arbitrary shape with any swimming velocity (Fig.~\ref{fig:fig1}). 

 The swimmer moves with translational and angular velocities $\bm{V}_\text{A}$ and $\bm{\Omega}_\text{A}$ (or in short $\bm{\mathsf{V}}_\text{A}=\left[\bm{V}_\text{A}~\bm{\Omega}_\text{A}\right]$), which result from a surface slip velocity profile $\bm{v}_\text{A}^\text{s}$ that is always tangential to the surface (i.e. $\bm{v}_\text{A}^\text{s}\cdot \bm{n}=0$, where $\bm{n}$ is the surface normal). We prove that in the space of all possible configurations of $\bm{v}_\text{A}^\text{s}$ that lead to the same $\bm{\mathsf{V}}_\text{A}$, the dissipation satisfies the inequality
\begin{equation}
\label{minimum}
P_\text{A}\geq \bm{\mathsf{V}}_\text{A}\cdot\left(\bm{\mathsf{R}}_\text{PS}^{-1} -\bm{\mathsf{R}}_\text{NS}^{-1}\right )^{-1}\cdot\bm{\mathsf{V}}_\text{A}.
\end{equation}
Here, $\bm{\mathsf{R}}_\text{PS}$ and  $\bm{\mathsf{R}}_\text{NS}$ are the rigid-body resistance tensors corresponding to the perfect-slip (PS) and no-slip (NS), and they map the translational and angular velocities to the net hydrodynamic forces and torques, namely $\bm{\mathsf{F}}_\text{PS}=\left[\bm{F}_\text{PS}~ \bm{L}_\text{PS}\right]$ and $\bm{\mathsf{F}}_\text{NS}=\left[\bm{F}_\text{NS}~ \bm{L}_\text{NS}\right]$. We thus demonstrate that the minimum dissipation at a given swimming speed---or the maximum swimming efficiency---of an active swimmer can be determined solely by the knowledge of the viscous resistance tensors of the same body with no-slip and perfect-slip boundary conditions. We use the theorem to propose a new expression for the {\it microswimmer efficiency} as 
\begin{align}
\label{eta-m-def}
\eta_\text{m}\equiv\frac{\bm{\mathsf{V}}_\text{A}\cdot\bm{\mathsf{R}}_\text{PS}\cdot\bm{\mathsf{V}}_\text{A}}{P_\text{A}} \le 1,
\end{align}
which compares the absolute minimum power needed to move the body with the same exact velocity (corresponding to dragging a perfect-slip body)
and the power expended by the active swimmer. We demonstrate that our proposed microswimmer efficiency is bounded by unity, unlike Lighthill efficiency.

We start the derivation by considering 
 an active swimmer of arbitrary shape, shown in Fig.~\ref{fig:fig1}. The motion of the swimmer is governed by the Stokes equations $\bm{\nabla}\cdot\bm{\sigma}=\bm{0}$ and $\bm{\nabla}\cdot\bm{v}={0}$ subject to boundary condition $\bm{v}(\bm{x}\in\mathcal{S})=\bm{V}_\text{A}+\bm{\Omega}_\text{A}\times \bm{x} +\bm{v}_\text{A}^\text{s}$. Here, $\bm{v}$ is the velocity field, $\bm{\sigma}=-p\bm{I}+2\mu\bm{E}$ is the stress field, $p$ is the pressure field, $\bm{E}=\left(\bm{\nabla}\bm{v}+\bm{\nabla}\bm{v}^\top\right)/2$ is the strain-rate tensor, $\mu$ is the fluid viscosity, $\mathcal{S}$ describes the surface of the particle, and $\bm{x}$ is the position vector.

The power dissipated due to any motion in a viscous fluid can be equivalently expressed either as a surface integral of the mechanical energy flux, $P=-\int_\mathcal{S}\text{d}S~\bm{n}\cdot\bm{\sigma}\cdot\bm{v}$, or as a volume integral of the density of dissipated power,
$P=2\mu \int_{\mathcal{V}}\text{d}V~\bm{E}:\bm{E}$. Here, $\mathcal V$ represents the volume of the surrounding fluid and the symbol~$:$ denotes a 2-fold contraction (i.e., $\bm{E}:\bm{E}=E_{ij}E_{ji}$).  The equivalence between the two expressions follows from the divergence theorem \cite{happel1983,guazzelli2009}. 
In the absence of any external force or torque (i.e., $\int_\mathcal{S}\text{d}S\bm{f}_\text{A}=\bm{0}$ and $\int_\mathcal{S}\text{d}S\bm{x}\times\bm{f}_\text{A}=\bm{0}$),
 the power dissipated by the swimmer then follows 
\begin{align}
P_\text{A}=-\int_\mathcal{S}\text{d}S \bm{f}_\text{A}\cdot\bm{v}^\text{s}_\text{A},
\end{align}
where $\bm{f}_\text{A}=\bm{n}\cdot\bm{\sigma}_\text{A}$ is the traction. Here, our main goal is to find a slip velocity profile $\bm{v}_\text{A}^\text{s}$ that minimizes $P_\text{A}$ for a swimmer of arbitrary shape, while keeping the swimming velocity $\bm{\mathsf{V}}_\text{A}$ constant. 

We start our derivation by first showing that among all flows around a body of a given shape, the flow that satisfies the perfect-slip boundary condition on the surface has the minimal dissipation. For this purpose we adapt the standard derivation of the Helmholtz minimum dissipation theorem (see \citet{guazzelli2009}), which states that the Stokes flow has the minimum dissipated power compared to any other flow that satisfies the same boundary condition on velocity. In our derivation, we compare the dissipation in the flow with the perfect-slip boundary with that in a flow with any other slip velocity.  We consider the motion of a passive perfect-slip body with tangential slip velocity $\bm{v}_\text{PS}^\text{s}$ and rigid-body motion $\bm{\mathsf{V}}_\text{PS}=[\bm{V}_\text{PS} ~\bm{\Omega}_\text{PS}]$. By definition, the tangential component of the traction in such a motion is zero, namely $\bm{f}_\text{PS}^{\parallel}=\left(\bm{I}-\bm{nn}\right)\cdot\bm{f}_\text{PS}=\bm{0}$. 
Any tangential perturbation in the slip profile of this motion (denoted by $'$) then alters the viscous dissipated power by
\begin{align}
\Delta P&=2\mu \int_\mathcal{V}\text{d}V\left[ \left(\bm{E}_\text{PS}+\bm{E}'\right):\left(\bm{E}_\text{PS}+\bm{E}'\right)-\bm{E}_\text{PS}:\bm{E}_\text{PS}\right]\nonumber\\
&=2\mu \int_\mathcal{V}\text{d}V \bm{E}':\bm{E}'+ 4\mu \int_\mathcal{V} \text{d}V\bm{E}':\bm{E}_\text{PS},
\end{align}
where $\Delta P$ is the change in the dissipation due to the perturbation in the slip profile and $\bm{E}'$ is the strain-rate tensor for the perturbation flow. Since both  $\bm{E}'$ and $\bm{E}_\text{PS}$ are traceless symmetric tensors, and $\bm{\nabla}\cdot\bm{v}'=0$, we have $2\mu \bm{E}':\bm{E}_\text{PS}=\bm{\nabla}\bm{v}':\bm{\sigma}_\text{PS}$, where $\bm{\sigma}_\text{PS}=-p_\text{PS}\bm{I}+2\mu\bm{E}_\text{PS}$ is the stress field for the perfect-slip flow and $p_\text{PS}$ is the corresponding pressure field. Using $\bm{\nabla}\cdot\bm{\sigma}_\text{PS}=\bm{0}$, one obtains $\bm{\nabla}\bm{v}':\bm{\sigma}_\text{PS}=\bm{\nabla}\cdot\left(\bm{v}'\cdot\bm{\sigma}_\text{PS}\right)$. Thence, by once again using the divergence theorem, we find $4\mu \int_\mathcal{V}\text{d}V \bm{E}':\bm{E}_\text{PS}=-2\int_\mathcal{S} \text{d}S \bm{f}_\text{PS}\cdot \bm{v}'=0$ (since $\bm{f}_\text{PS}^{\parallel}=\bm{0}$ and $\bm{n}\cdot \bm{v}'=0$). Now, noting that $2\mu \int_\mathcal{V} \text{d}V \bm{E}':\bm{E}'$ is positive-definite, we can conclude that the minimum dissipated power can be only achieved when $\bm{E}'=\bm{0}$, thereby indicating that the optimal slip velocity profile is that of a perfect-slip body. We can alternatively state that the dissipation of any motion with rigid-body velocity $\bm{\mathsf{V}}_\text{PS}$ is more than (or equal to) that of a perfect-slip body. For a perfect-slip body, the dissipated power is found $P_\text{PS}=-\int_{\mathcal{S}}\text{d}S \bm{f}_\text{PS}\cdot\bm{v}_\text{PS}=\bm{\mathsf{V}}_\text{PS}\cdot\bm{\mathsf{R}}_\text{PS}\cdot\bm{\mathsf{V}}_\text{PS}$.
Thus, we have proven
\begin{align}
\label{inequality}
P\geq \bm{\mathsf{V}}\cdot\bm{\mathsf{R}}_\text{PS}\cdot\bm{\mathsf{V}},
\end{align}
where $P$ is the dissipated power in any flow with rigid-body motion $\bm{\mathsf{V}}$. We note that there is an interesting difference between the implications of the Helmholtz theorem for no-slip and perfect-slip bodies. For no-slip bodies, the Helmholtz theorem implies that adding extra volume to the body, such that the old shape is fully contained in the new one, will always increase its drag coefficient \cite{Hill.Power1956}. This is because the flow around the enlarged body can be viewed as a non-optimal solution to the original problem. For perfect-slip bodies, however, no such statement is possible. For example, a perfect-slip spheroid reduces its drag coefficient upon elongation in the direction of motion while keeping the equatorial radius constant \cite{supp_efficiency2}.

\begin{figure}
\centering
\includegraphics[width=\columnwidth]{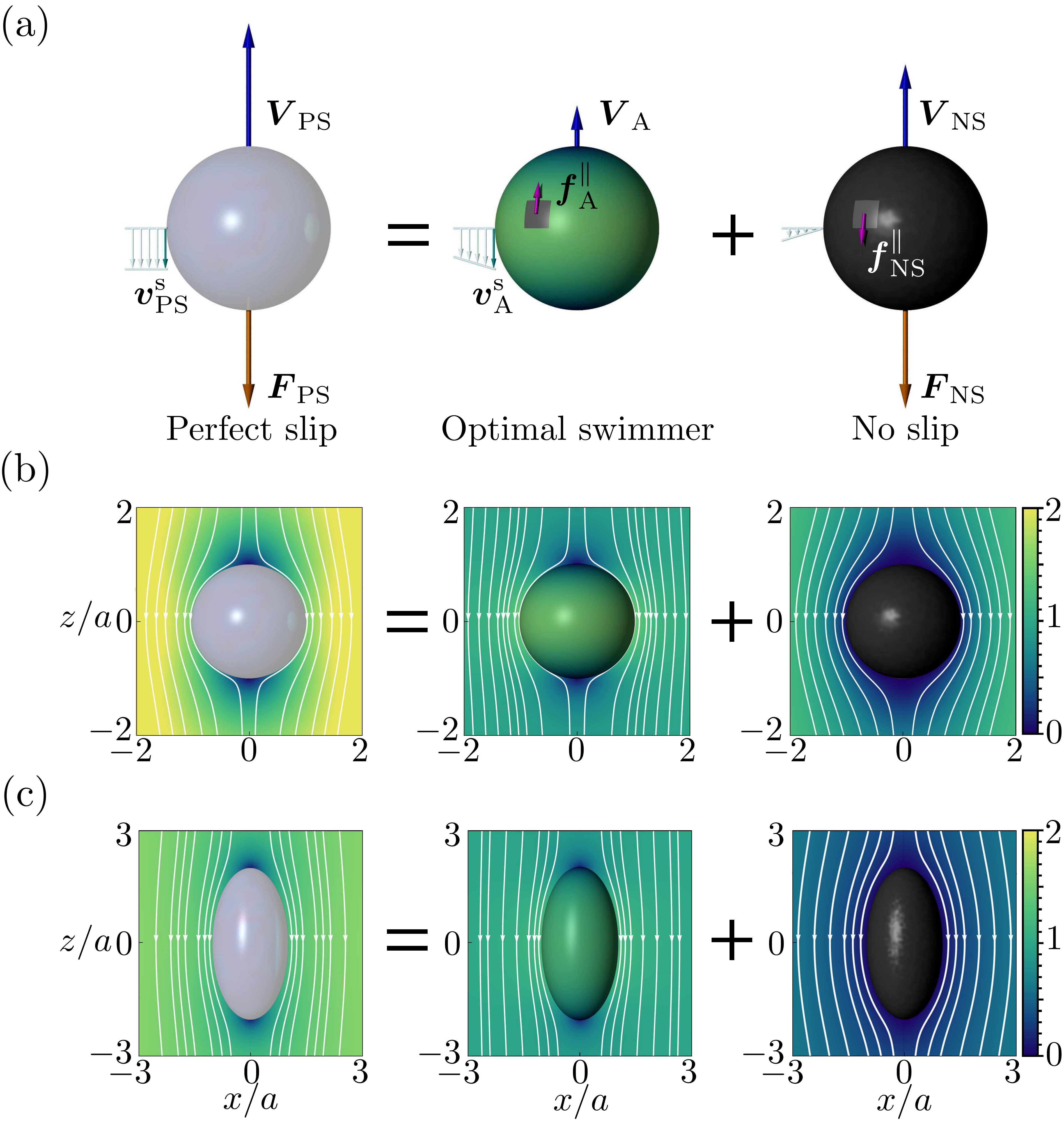}
\caption{(a) Schematic of the superposition. A perfect-slip body is represented as a superposition of the optimal active swimmer and a no-slip body. The cyan-colored arrows schematically show the velocity profile near the surface of the swimmer (in the co-moving frame). 
Magenta-colored arrows show the tangential traction force density. This superposition is shown for two examples of (b) a sphere of radius $a$, and (c) a prolate spheroid with aspect ratio $b/a=2$. The white arrows show the streamlines and the colors indicate the value of the velocity scaled by the swimming speed of the active particle. All the figures are in the co-moving Cartesian $x$-$z$ frame set at the center of the particle.}
  \label{fig:fig2}
\end{figure}

We now need to show how the inequality given in Eq.~\eqref{inequality} sets an absolute lower bound for the dissipated power of an active swimmer of the same shape with velocity $\bm{\mathsf{V}}_\text{A}$. Let us consider the motion of a body that is a linear superposition of an active swimmer with velocity $\bm{\mathsf{V}}_\text{A}$ and a no-slip body of the same shape with velocity $\bm{\mathsf{V}}_\text{NS}$. From \eqref{inequality}, we have a lower bound on the dissipated power in the superposition system
\begin{equation}
\label{inequality1}
  P_\text{A+NS}\geq (\bm{\mathsf{V}}_\text{A}+\bm{\mathsf{V}}_\text{NS}  ) \cdot \bm{\mathsf{R}}_\text{PS} \cdot(\bm{\mathsf{V}}_\text{A}+\bm{\mathsf{V}}_\text{NS}).
\end{equation}


On the other hand, the dissipated power can be directly expressed as $P_\text{A+NS}=-\int_\mathcal{S}\text{d}S  (\bm{f}_\text{A}+\bm{f}_\text{NS}) \cdot\bm{v}_\text{A+NS}$, where $\bm{v}_\text{A+NS}=\bm{v}^\text{s}_\text{A}+\bm{V}_\text{A}+\bm{V}_\text{NS}+\left(\bm{\Omega}_\text{A}+\bm{\Omega}_\text{NS}\right)\times \bm{x}$ at the boundary. This dissipated power then evaluates to

\begin{align}
  \label{psuperpos}
P_\text{A+NS}&=-\int_\mathcal{S}\text{d}S  (\bm{f}_\text{A}+\bm{f}_\text{NS}) \cdot\bm{v}^\text{s}_\text{A}- \bm{\mathsf{F}}_\text{NS}\cdot( \bm{\mathsf{V}}_\text{A} + \bm{\mathsf{V}}_\text{NS}) .
\end{align} 

Now we can employ the Lorentz reciprocal theorem to connect the active swimming problem to the no-slip one \cite{Lorentz1896,happel1983,stone1996,elfring2017,nasouri2018,masoud2019}. Since the flow field in the active problem is force- and torque-free, we find
$\int_\mathcal{S}\text{d}S \bm{f}_\text{NS}\cdot\bm{v}^\text{s}_\text{A}=-\bm{\mathsf{F}}_\text{NS}\cdot\bm{\mathsf{V}}_\text{A}$, which simplifies Eq.~(\ref{psuperpos}) to
\begin{equation}
  \label{poweradditivity}
P_\text{A+NS}=P_\text{A}-\bm{\mathsf{F}}_\text{NS}\cdot\bm{\mathsf{V}}_\text{NS},
\end{equation}
where $P_\text{A}$ and $-\bm{\mathsf{F}}_\text{NS}\cdot\bm{\mathsf{V}}_\text{NS}$ represent the dissipated power of the active and the no-slip particles, respectively. Thus, for the superposition of a force- and torque-free active particle and a no-slip body with arbitrary velocity, the dissipation rate is the sum of the two corresponding individual dissipation rate contributions.

The lower bound on the dissipation rate---given by Eqs.~\eqref{inequality1} and \eqref{poweradditivity}---is valid for a superposition of an active swimmer with any slip profile 
$\bm{v}_\text{A}^\text{s}$ (resulting in the swimming velocity $\bm{\mathsf{V}}_\text{A}$) and a no-slip body with any velocity $\bm{\mathsf{V}}_\text{NS}$. The equality (i.e., the minimum dissipation for the swimmer) is fulfilled if and only if 
the superposition represents a flow around a perfect-slip body.  Among all $\bm{\mathsf{V}}_\text{NS}$, we therefore obtain the strictest bound on the dissipation when the velocities fulfill the conditions of the superposition described in Fig.~\ref{fig:fig2}(a). The velocities then must follow $\bm{\mathsf{V}}_\text{PS} = \bm{\mathsf{V}}_\text{A}+\bm{\mathsf{V}}_\text{NS}$, and the force balance of the superposition dictates ${\bm{\mathsf{R}}}_\text{PS}\cdot{\bm{\mathsf{V}}}_\text{PS}={\bm{\mathsf{R}}}_\text{NS}\cdot{\bm{\mathsf{V}}}_\text{NS}$ (since ${\bm{\mathsf{F}}}_\text{PS}=\bm{\mathsf{F}}_\text{NS}$).
The velocities can then be expressed in terms of $\bm{\mathsf{V}}_\text{A}$ as

\begin{align}
\label{PS}
{\bm{\mathsf{V}}}_\text{PS}&=\left(\bm{\mathsf{I}} -\bm{\mathsf{R}}_\text{NS}^{-1} \cdot\bm{\mathsf{R}}_\text{PS}\right)^{-1}\cdot{\bm{\mathsf{V}}}_\text{A},\\
\label{NS}
{\bm{\mathsf{V}}}_\text{NS}&=\left(\bm{\mathsf{R}}_\text{PS}^{-1} \cdot\bm{\mathsf{R}}_\text{NS}-\bm{\mathsf{I}}\right )^{-1}\cdot{\bm{\mathsf{V}}}_\text{A},
\end{align}
where $\bm{\mathsf{I}}$ is the six-dimensional identity tensor. Inserting these velocities in Eqs.~\eqref{inequality1} and \eqref{poweradditivity} yields our main result that is presented in Eq.~\eqref{minimum}.
Equation \eqref{minimum} shows that the minimum dissipated power by a swimmer can be obtained explicitly by evaluating the resistance tensors of two passive problems: the perfect-slip body and the no-slip body motion of the same geometry. This result is general and holds for any swimmer of any arbitrary shape (the simplified version for axisymmetric swimmers is presented in Supplemental Material \cite{supp_efficiency2}). The optimization problem is remarkably reduced down to finding the resistance tensor of two passive systems in the same domain. 

As discussed, the power dissipated in active swimming reaches its lower bound when the superposition of the active and no-slip cases conspires to exactly recreate the perfect-slip motion.  Note that an unlimited variety of slip profiles can lead to the same swimming velocity, but only the superposition of the optimal swimmer and the no-slip body can amount to the flow field of a perfect-slip body. In other words, such a decomposition of the perfect-slip body uniquely demands $\bm{v}_\text{A}^\text{s}=\bm{v}_\text{PS}^\text{s}$.  From this, one can claim that the optimal slip velocity profile for any swimmer with swimming velocity $\bm{\mathsf{V}}_\text{A}$ is identical to that of a perfect-slip body moving with a velocity given in Eq.~\eqref{PS}. Furthermore, since for the perfect-slip body $\bm{{f}}_\text{PS}^{\parallel}=\bm{0}$, the superposition also necessitates $\bm{f}_\text{A}^{\parallel}=-\bm{f}_\text{NS}^{\parallel}$. Thus, we can similarly claim that the force density of an optimal swimmer is the negative of that induced by a passive motion of a no-slip body with the velocity given in \eqref{NS}.

\begin{figure}
\centering\includegraphics[width=0.85\columnwidth]{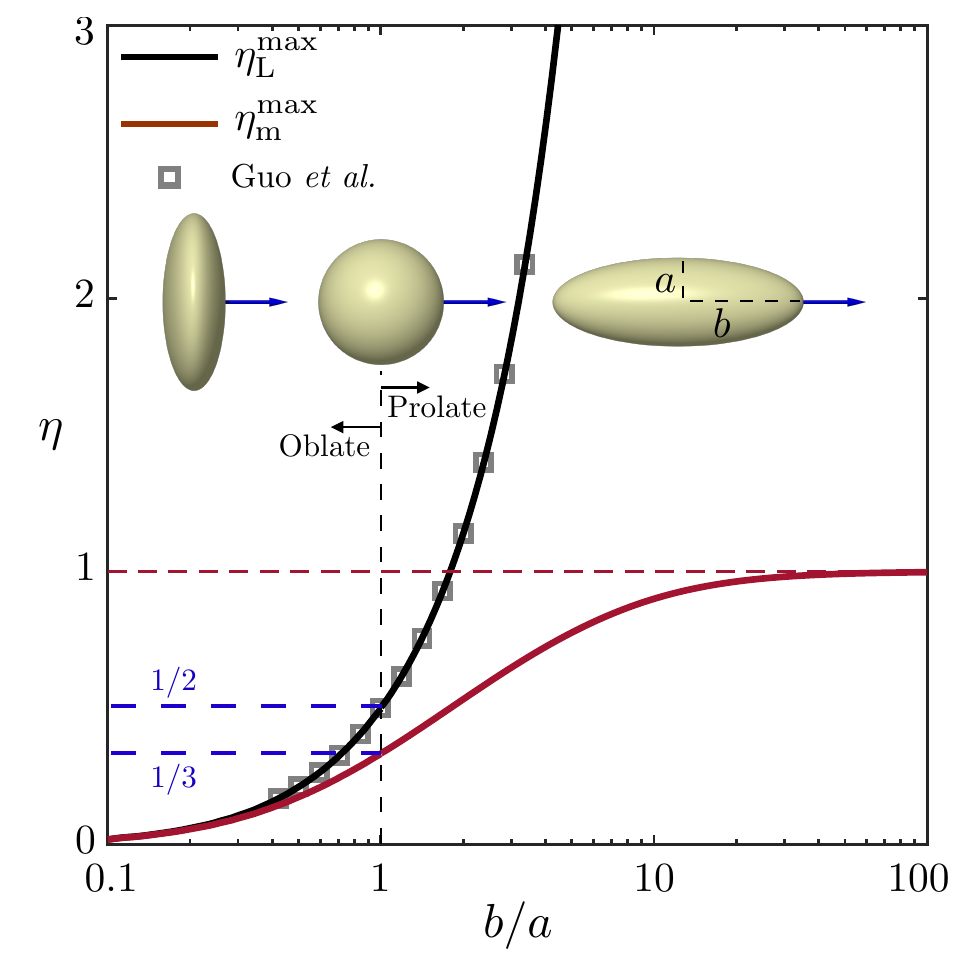}
\caption{Maximum Lighthill efficiency $\eta_\text{L}^\text{max}$ (black line) of spheroidal swimmers as a function of the aspect ratio $b/a$, obtained using our theorem. The Lighthill efficiency diverges for prolate spheroids as $b/a$ increases. The red line shows our proposed microswimmer efficiency $\eta_\text{m}^\text{max}$, which is bounded by unity. For a sphere ($b/a=1$), our theorem recovers the well-documented maximum Lighthill efficiency of $1/2$ \cite{michelin2010}, which corresponds to $1/3$ in the case of our proposed microswimmer efficiency. The grey squares show optimal efficiencies obtained by \citet{guo2020} using BEM and numerical optimization.}
\label{fig:fig3}
\end{figure}

The expression for the minimum power can also be used to find the maximum Lighthill efficiency. 
Recalling that the total power needed for dragging a no-slip body with velocity $\bm{\mathsf{V}}_\text{A}$ is $\bm{\mathsf{V}}_\text{A}\cdot \bm{\mathsf{R}}_\text{NS}\cdot\bm{\mathsf{V}}_\text{A}$, we find
\begin{align}
\label{lighthill}
\eta_\text{L}\le \frac{\bm{\mathsf{V}}_\text{A}\cdot\bm{\mathsf{R}}_\text{NS}\cdot\bm{\mathsf{V}}_\text{A}}{\bm{\mathsf{V}}_\text{A}\cdot\left(\bm{\mathsf{R}}_\text{PS}^{-1} -\bm{\mathsf{R}}_\text{NS}^{-1}\right )^{-1}\cdot\bm{\mathsf{V}}_\text{A}}.
\end{align}
Expression \eqref{lighthill} is also general and sets the upper bound for the efficiency of a swimmer with a given shape and swimming velocity. If the motion is axisymmetric with no rotation, the limit on efficiency takes the very simple form 
\begin{align}
  \label{lighthillAS}
\eta_\text{L}\le \frac{R_\text{NS}}{R_\text{PS}}-1,
\end{align}
where $R_\text{NS}$ and $R_\text{PS}$ are now the scalar drag coefficients of the no-slip and perfect-slip passive bodies. For instance, for a spherical swimmer (i.e., a squirmer \cite{lighthill1952,blake1971,pak2014}), we have $R_\text{NS}=6\pi \mu a$ and $R_\text{PS}=4\pi \mu a$, where $a$ is the radius \cite{happel1983}. This gives $\eta^\text{max}_\text{L}=1/2$, a result which was also found by \citet{michelin2010}. From the known solutions for the flow around a bubble we know that the slip velocity of the optimal spherical swimmer is $v^\text{s}_\text{A}=\frac 32 V_\text{A} \sin \theta$ with $\theta$ being the polar angle, also in agreement with Refs.\ \cite{Blake1973,michelin2010}. The force density, which is identical to the force on a no-slip sphere translating with velocity $(R_\text{NS}/R_\text{PS} -1)^{-1}V_\text{A}$, is then simply found as $f^\parallel_\text{A}=3 \mu (V_\text{A}/a)  \sin \theta$. The flow decomposition for an optimal spherical swimmer is shown in  Fig.~\ref{fig:fig2}(b).

To highlight the strength of expression \eqref{lighthillAS}, we may also use it to evaluate the maximum efficiency of an axisymmetric spheroidal swimmer. The exact expression for the flow field and the drag coefficient of a no-slip spheroidal body is well documented \cite{happel1983}, and we can similarly find the flow field and the drag coefficient for the perfect-slip body using an exact approach \cite{supp_efficiency2} [see Fig.~\ref{fig:fig2}(c) for an example]. By using these results, we can evaluate the maximum Lighthill efficiency for any value of the spheroid aspect ratio $b/a$. As shown in Fig.~\ref{fig:fig3}, our results precisely match the recent computational data for optimal swimming of spheroidal particles obtained with the boundary element method (BEM) and numerical optimization by \citet{guo2020}. These results provide a fully independent validation of our theorem. 

Our simple expression for the maximum efficiency also shows how the Lighthill efficiency diverges once the aspect ratio becomes increasingly large $b/a\rightarrow \infty$ (spheroid transforms to a needle) \cite{leshansky2007}. In that case, ${R}_\text{NS}\propto b/\log(b/a) \rightarrow \infty$ while ${R}_\text{PS}\propto a^2/b \to 0$, resulting in $\eta_\text{L}^\text{max}\propto (b/a)^2/\log(b/a)\rightarrow \infty$. To resolve this, we propose an alternative microswimmer efficiency via Eq.~\eqref{eta-m-def}, which yields
\begin{align}
\label{lighthillAlt}
\eta_\text{m}\le \frac{\bm{\mathsf{V}}_\text{A}\cdot\bm{\mathsf{R}}_\text{PS}\cdot\bm{\mathsf{V}}_\text{A}}{\bm{\mathsf{V}}_\text{A}\cdot\left(\bm{\mathsf{R}}_\text{PS}^{-1} -\bm{\mathsf{R}}_\text{NS}^{-1}\right )^{-1}\cdot\bm{\mathsf{V}}_\text{A}},
\end{align}
and consequently
\begin{align}
\eta_\text{m} \le 1-\frac{R_\text{PS}}{R_\text{NS}},
\end{align}
for axisymmetric bodies. Using this definition, for a spherical swimmer $\eta^\text{max}_\text{m}=1/3$, and for a needle $\eta^\text{max}_\text{m}\rightarrow 1$; see Fig.~\ref{fig:fig3}. We have demonstrated that the motion of a perfect-slip body has the minimum dissipation among all other types of motion, and therefore, unlike the Lighthill efficiency, $\eta_\text{m}$ is clearly bounded by unity. 

In conclusion, we were able to express the minimum dissipation needed by an active swimmer using only two rigid-body resistance tensors of bodies with the same shape: a no-slip boundary condition in the first case and a perfect-slip boundary in the second. We showed that the surface slip velocities and forces that reach this minimal dissipation correspond to the flow around a perfect-slip body and the tangential force on the no-slip body. We have thus reduced a complex optimization problem to the calculation of two resistance tensors, for which numerous analytical solutions and numerical methods are available.

In this study, we found a general theorem on the minimum dissipation of surface-driven microswimmers. An outstanding challenge will be to generalize this theorem to swimmers that propel themselves by changing their shapes \cite{najafi2004,jalali2014,mirzakhanloo2018}.

We thank Evelyn Tang for comments on the manuscript. This work has been supported by the Max Planck Society. A.V. acknowledges support from the Slovenian Research Agency (grant no. P1-0099).

\bibliography{reference}

\end{document}


\title{Supplemental Material -- Minimum Dissipation Theorem for Microswimmers}
\author{Babak Nasouri}
\affiliation{Max Planck Institute for Dynamics and Self-Organization (MPIDS), 37077 G\"ottingen, Germany}
\author{Andrej Vilfan}
\email{andrej.vilfan@ds.mpg.de}
\affiliation{Max Planck Institute for Dynamics and Self-Organization (MPIDS), 37077 G\"ottingen, Germany}
\affiliation{Jo\v{z}ef Stefan Institute, 1000 Ljubljana, Slovenia}
\author{Ramin Golestanian}
\affiliation{Max Planck Institute for Dynamics and Self-Organization (MPIDS), 37077 G\"ottingen, Germany}
\affiliation{Rudolf Peierls Centre for Theoretical Physics, University of Oxford, Oxford OX1 3PU, United Kingdom}
\date{\today}

\maketitle
\section{Axisymmetric swimmers}
The theorem and equations given in the main article are general and valid for any swimmer of any arbitrary shape. Here we provide a summary of the theorem with simplified expressions for axisymmetric swimmers. To this end, we consider an axisymmetric swimmer, swimming with speed $V_\text{A}$. Defining $R_\text{PS}$ and $R_\text{NS}$ as the drag coefficients of the perfect-slip and no-slip bodies of the same shape along the same axis of symmetry, the theorem states that the dissipated power by the swimmer is bounded by
\begin{align}
P_A\geq \frac{V_\text{A}^2}{\frac{1}{R_\text{PS}}-\frac{1}{R_\text{NS}}}.
\end{align}
The theorem also states that the optimal swimmer has a slip profile identical to that of the perfect-slip body translating with speed 
\begin{align}
V_\text{PS}=\frac{V_\text{A}}{1-R_\text{PS}/R_\text{NS}},
\end{align}
and a force density which is the negative of that induced by the motion of the no-slip body translating with speed 
\begin{align}
V_\text{NS}=\frac{V_\text{A}}{R_\text{NS}/R_\text{PS}-1}.
\end{align}
Using the theorem, we find the upper limit for the Lighthill efficiency as 
\begin{align}
\eta_\text{L}\equiv\frac{R_\text{NS} V_\text{A}^2}{P_A}\leq  \frac{R_\text{NS}}{R_\text{PS}} -1,
\end{align}
and define an alternative microswimmer efficiency as
\begin{align}
\eta_\text{m}\equiv\frac{R_\text{PS} V_\text{A}^2}{P_A}\leq  1-\frac{R_\text{PS}}{R_\text{NS}} ,
\end{align}
which unlike the Lighthill efficiency, is always bounded by unity.

\section{Axisymmetric motion of a perfect-slip spheroidal particle}
Here, we show the details of our calculations for finding the flow field and the drag coefficient of a perfect-slip spheroidal particle, needed for finding the optimal slip profile of a spheroidal microswimmer (Figs.~2 and 3 in the main article).

\begin{figure}
\begin{center}
\includegraphics[width=0.4\columnwidth]{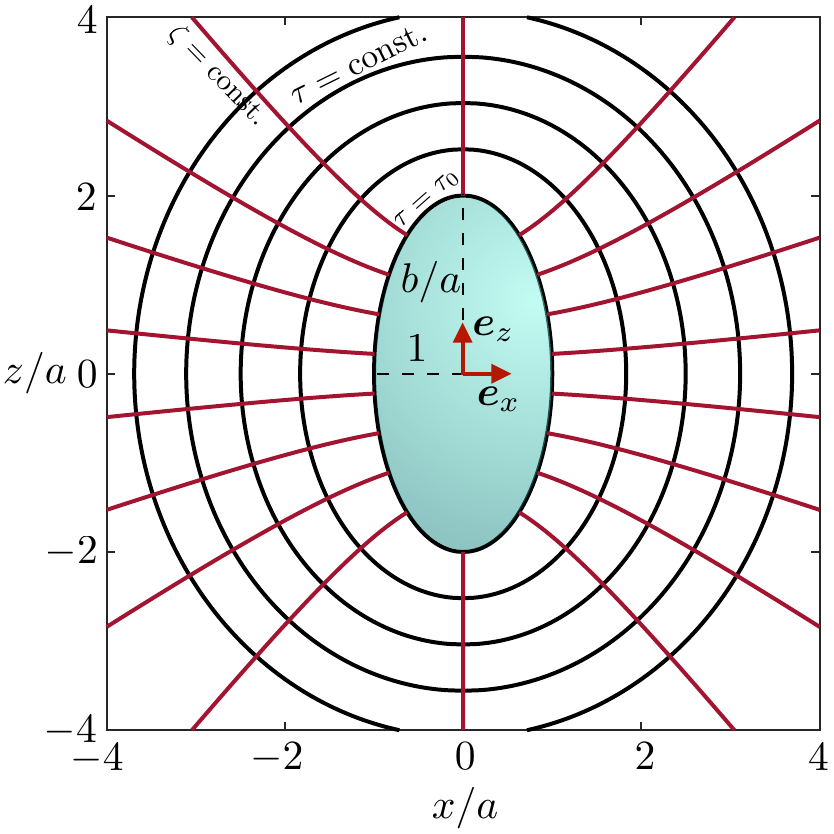}
\caption{Schematic of a prolate spheroid with aspect ratio $b/a=2$, in the prolate spheroidal coordinate system $(\tau,\zeta,\phi)$. The value of $\tau$ remains constant along the black lines, and lines of constant $\zeta$ are in red.}
\label{fig}
\end{center}
\end{figure}

We begin by considering a prolate spheroidal particle with semi-axes $b$ and $a$ ($b>a$), aligned along axes $\bm{e}_z$ and $\bm{e}_x$, as shown in Fig.~\ref{fig}. We assume the motion is axisymmetric with translational velocity $\bm{V}=V\bm{e}_z$. The flow field due to the motion of the particle is governed by the Stokes equations
\begin{align}
\label{stokes}
\bm{\nabla}\cdot\bm{\sigma}(\bm{x})=\bm{0},\quad\bm{\nabla}\cdot\bm{v}(\bm{x})=0,
\end{align}
where $\bm{\sigma}=-p\bm{I} + \mu \left(\bm{\nabla}\bm{v}+\bm{\nabla}\bm{v}^\top\right)$ is the stress field, $\bm{v}$ and ${p}$ are the flow field and pressure field, $\bm{x}$ is the position vector, and $\mu$ is the fluid viscosity. 
In the co-moving frame, the boundary conditions for this motion take the form
\begin{align}
\label{inf}
&\bm{v}(\bm{x})=-\bm{V},\quad|\bm{x}|\rightarrow\infty\\ 
\label{surf}
&\bm{n}\cdot\bm{v}(\bm{x})={0},\quad\bm{x}\in\mathcal{S}
\end{align}
where $\mathcal{S}$ describes the surface of the particle, and $\bm{n}$ is a unit vector normal to $\mathcal{S}$. Here, Eq. \eqref{inf} indicates that the effect of the particle must vanish when $|\bm{x}|\rightarrow\infty$, and Eq. \eqref{surf} imposes the impermeability condition at the surface. By definition, a perfect-slip body cannot have any tangential traction 
\begin{align}
\label{slip}
\left(\bm{I}-\bm{nn}\right)\cdot(\bm{n}\cdot\bm{\sigma})=\bm{0}.
\end{align}

To go further, it is convenient to take a prolate spheroidal coordinate system set at the center of the particle. As it has been well discussed for no-slip bodies \cite{happel1983,dassios1994}, droplets \cite{deo2003}, and active particles \cite{pohnl2020}, using this coordinate system would allow us to find the stream function in terms of an infinite series. Here, we closely follow the methodology described by \citet{dassios1994} (or its adaptation by \citet{pohnl2020}) and define
\begin{align}
x=c\sqrt{\tau^2-1}\sqrt{1-\zeta^2}\cos\phi,\quad y=c\sqrt{\tau^2-1}\sqrt{1-\zeta^2}\sin\phi,\quad z= c \tau\zeta,
\end{align}
where $0<\tau<\infty$, $-1<\zeta<1$, $0<\phi<2\pi$, and $c=\sqrt{b^2-a^2}$. Under these definitions, the surface of the spheroid can be simply described by $\tau=\tau_0=b/c$, and also we have $\bm{n}=\bm{e}_\tau|_{\tau=\tau_0}$.

Since the motion is axisymmetric, we can alternatively write down the Stokes equations given in \eqref{stokes} using the stream function as 
\begin{align}
\label{stream}
\nabla^4\psi=0. 
\end{align}
Defining $\bm{v}=v_\tau\bm{e}_\tau+v_\zeta\bm{e}_\zeta $, the stream function is connected to components of the velocity as
\begin{align}
v_\tau=\frac{1}{c^2\sqrt{\tau^2-\zeta^2}\sqrt{\tau^2-1}}\frac{\partial\psi}{\partial\zeta},\quad v_\zeta=\frac{-1}{c^2 \sqrt{\tau^2-\zeta^2}\sqrt{1-\zeta^2}}\frac{\partial\psi}{\partial\tau}. 
\end{align}
As shown by \citet{dassios1994}, the general solution to the bi-harmonic equation given in \eqref{stream} can be found in terms of Gegenbauer functions of the first kind ($G$) and second kind ($H$). Noting that in our problem, the stream function must be finite everywhere, the general solution takes the form
\begin{align}
\label{general}
\psi(\tau,\zeta)=\sum_{n=2}^\infty g_n(\tau) G_n(\zeta),
\end{align}
where
\begin{align}
g_2(\tau)=B_2 G_1(\tau) + A G_2(\tau) + D_2 H_2(\tau)+B_4 H_4(\tau), 
\end{align}
and for any $n\geq4$
\begin{align}
g_n(\tau)=B_n H_{n-2}(\tau) + D_n H_n(\tau) +B_{n+2} H_{n+2}(\tau),
\end{align}
where $B_n$, $D_n$, and $A$ are constant coefficients that need to be determined via applying the boundary conditions. Note that here we only need to keep the even terms of the general solution $n\in\{2,4,6,\cdots\}$. This is because the stream function must be symmetric with respect to $\zeta$, and thus should not change via the transformation $\zeta\rightarrow-\zeta$. 

To determine the coefficients, we need to truncate the solution at some order, say $N$, and assume the coefficients of higher orders are identically zero. We then have $N+1$ unknowns which require $N+1$ equations. We begin by the boundary condition given in \eqref{inf}. We find $\psi(|\bm{x}|\rightarrow\infty)=\frac{1}{2}Vc^2\left(\tau^2 -1\right)\left(1-\zeta^2\right)$. Recalling that $G_2(x)=\frac{1}{2}\left(1-x^2\right)$, we simply arrive at 
\begin{align}
\label{c2}
A=-2V c^2. 
\end{align} 
The boundary condition given in \eqref{surf}, indicates that at $\tau=\tau_0$ we have $v_\tau=0$. This implies $\partial\psi/\partial\zeta(\tau=\tau_0)=0$. But since $\psi(\tau=\tau_0,\zeta=\pm1)=0$, one can then claim $\psi(\tau=\tau_0)=0$ leading to our next set of equations
\begin{align}
\label{no-penetration}
g_n(\tau_0)=0.
\end{align} 
To close the system of equations, we now need to use the perfect-slip boundary condition expressed in \eqref{slip}. In the prolate spheroidal coordinate system, this condition simplifies to
\begin{align}
\label{slip2}
\frac{2\tau_0}{\tau_0^2-\zeta^2}\frac{\partial\psi}{\partial \tau}|_{\tau=\tau_0}=\frac{\partial^2\psi}{\partial \tau^2}|_{\tau=\tau_0}. 
\end{align}
\begin{figure}
\begin{center}
\includegraphics[width=0.40\columnwidth]{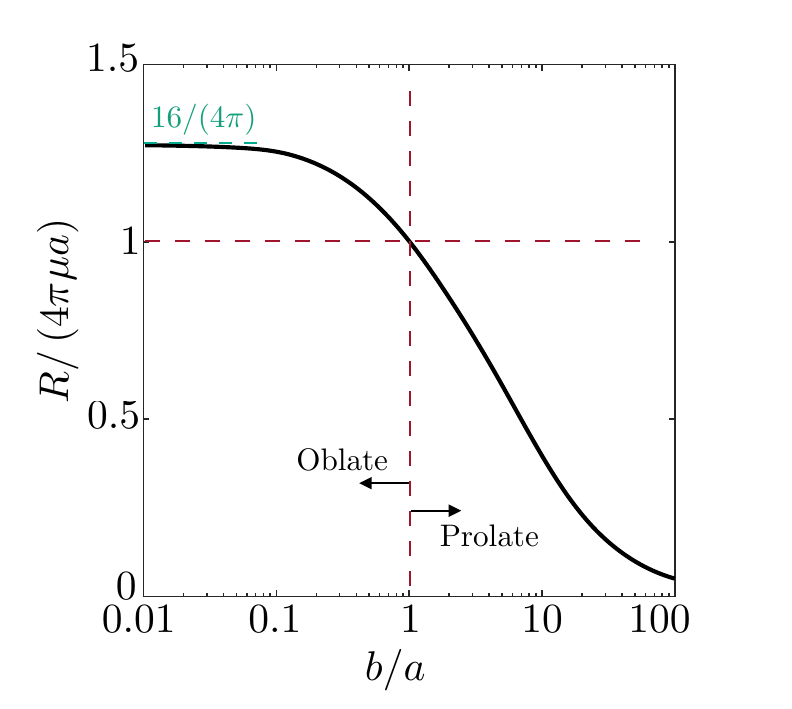}
\caption{The drag coefficient of perfect-slip spheroidal particles with semi-axes $b$ and $a$. }
\label{drag}
\end{center}
\end{figure}
It is now convenient to make use of the following recursive relations:
\begin{align}
\zeta^2 G_2(\zeta)&=\frac{1}{5}G_2(\zeta) + \frac{4}{5} G_4(\zeta),\nonumber \\
\zeta^2 G_n(\zeta)&=\alpha_nG_{n-2}(\zeta) + \gamma_n G_n(\zeta)+\beta_n G_{n+2}(\zeta),\quad \text{for~}{n\geq4}
\end{align}
where 
\begin{align}
\alpha_n=\frac{(n-3)(n-2)}{(2n-3)(2n-1)},\quad \beta_n=\frac{(n+1)(n+2)}{(2n-1)(2n+1)},\quad \gamma_n=\frac{2n^2 -2n -3}{(2n+1)(2n-3)}.
\end{align}
Now by substituting \eqref{general} into \eqref{slip2}, and using these recursive relations, the perfect-slip boundary condition finally reduces to
\begin{align}
\label{slip3}
2\tau_0g'_2&=\tau_0^2g''_2 - \frac{1}{5}g''_2 - \alpha_4 g''_4,\nonumber\\
2\tau_0g'_4&=\tau_0^2g''_4 -\gamma_n g''_4 - \frac{4}{5}g''_2-\alpha_6g''_6,\nonumber\\
2\tau_0g'_n&=\tau_0^2g''_n - \gamma_n g''_n - \beta_{n-2} g''_{n-2} -\alpha_{n+2}g''_{n+2},\quad \text{for~} n\geq6 
\end{align}
where $g'_n=\frac{\text{d} g_n}{\text{d} \tau}|_{\tau=\tau_0}$ and $g''_n=\frac{\text{d}^2 g_n}{\text{d} \tau^2}|_{\tau=\tau_0}$. Now, by using Eqs. \eqref{c2}, \eqref{no-penetration}, and \eqref{slip3}, we can construct a linear system of equations which can be solved to find the coefficients $B_n$ and $D_n$. By using these coefficients, we can determine the stream function and thus the flow field surrounding a perfect-slip prolate spheroidal particle. For the drag coefficient, $R$, one can use the results of \citet{payne1960} for an axisymmetric body which in our case reads $R=4 \pi c B_2$. We can then simply calculate the drag coefficient for different aspect ratios, as shown in Fig.~\ref{drag}.
All of the expressions derived here can be simply mapped to oblate spheroidal coordinates, which then can be used to resolve the flow field surrounding the motion of a perfect-slip oblate spheroidal particle. As discussed by \citet{happel1983}, one can use the transformation $\lambda=i \tau$ and ${c}'=-i c$, and define $\lambda_0=b/c'$. By using this mapping, we can determine the drag coefficient of the oblate spheroids, also shown in Fig.~\ref{drag}. As a perfect-slip oblate spheroids transforms to a circular disc ($b\rightarrow 0$), one can see that the drag coefficient asymptotes to the one of a no-slip circular disc which is $16\mu a$ \cite{happel1983}.
We finally note that for the results reported in the main article, we set $N=20$, as further increasing the number of terms did not result in any changes in the values. 

\bibliography{reference}